\documentstyle[12pt]{article}   
\def\lapproxeq{\lower .7ex\hbox{$\;\stackrel{\textstyle <}{\sim}\;$}}
\def\gapproxeq{\lower .7ex\hbox{$\;\stackrel{\textstyle >}{\sim}\;$}}

\newcommand{\be}{\begin{equation}}
\newcommand{\ee}{\end{equation}}

\begin{document}           

\newfont{\sevenrm}{cmr7}
\newfont{\teni}{cmmi10} 
\newfont{\seveni}{cmmi7} 
\newfont{\sevensy}{cmsy7} 
\newfont{\fiverm}{cmr5} 
\newfont{\fivei}{cmmi5} 
\newfont{\fivesy}{cmsy5} 

\def\tenpoint{\normalbaselineskip=12pt 
\abovedisplayskip 12pt plus 3pt minus 9pt 
\belowdisplayskip 12pt plus 3pt minus 9pt 
\abovedisplayshortskip 0pt plus 3pt 
\belowdisplayshortskip 7pt plus 3pt minus 4pt 
\smallskipamount=3pt plus1pt minus1pt 
\medskipamount=6pt plus2pt minus2pt 
\bigskipamount=12pt plus4pt minus 4pt 
\def\rm{\fam0\tenrm}          \def\it{\fam\itfam\tenit}%
\def\sl{\fam\slfam\tensl}      \def\bf{\fam\bffam\tenbf}%
\def\smc{\tensmc}               \def\mit{\fam 1}%
\def\cal{\fam 2}%
\textfont0=\tenrm      \scriptfont0=\sevenrm    \scriptscriptfont0=\fiverm 
\textfont1=\teni       \scriptfont1=\seveni     \scriptscriptfont1=\fivei
\textfont2=\tensy      \scriptfont2=\sevensy    \scriptscriptfont2=\fivesy
\textfont3=\tenex      \scriptfont3=\tenex      \scriptscriptfont3=\tenex 
\normalbaselines\rm}

\titlepage 
\begin{flushright} 
Durham DTP/94/56 \\ CERN-TH.7357/94           
\end{flushright} 

\begin{center}
\vspace*{2cm}
{\large{\bf Dijet production at HERA as a probe of BFKL dynamics}} 
\end{center} 

\vspace*{1cm} 
\begin{center} 
A.J.\ Askew$^1$, D.\ Graudenz$^2$, J.\ Kwiecinski$^3$ and A.D.\ Martin$^1$ 
\end{center} 
\vspace*{ 2 cm}
\begin{center}

\noindent $^1$ Department of Physics, University of Durham, Durham 
\\
\ $^2$ Theoretical Physics Division, CERN, CH-1211 Geneva 23\\ 
\ $^3$ Henryk Niewodniczanski Inst.\ Nuclear Physics, Krakow   

 \end{center}  
\vspace*{3cm} 

\begin{abstract}   
We calculate the rate for the deep-inelastic electroproduction of dijets at 
HERA.  We study the weakening of the azimuthal (back-to-back) correlation 
between the jets, as $x$ decreases, to see whether it can be used to 
identify BFKL dynamics from conventional fixed-order QCD effects.      
We show how this may give information on the transverse momentum 
($k_T)$ dependence
of the gluon distribution of the proton.
\end{abstract} 

\vfill

\begin{flushleft} 
CERN-TH.7357/94  \\ July 1994         
\end{flushleft}

\newpage

The production of dijets at the HERA electron-proton collider offers an 
excellent opportunity to study the properties of the gluon distribution of the
proton at small $x$. At lowest order, dijets\footnote{As is 
customary, we use the term \lq\lq dijet" to refer to two jets produced in 
addition to the jet formed by the remnants of the proton.} are produced by the 
emission of a \lq\lq hard\rq\rq\ gluon from the initial or final state of
the struck quark (the QCD Compton process $\gamma q \rightarrow g q$) or
by photon-gluon fusion ($\gamma g \rightarrow q \overline{q}$). The dijet
events of particular interest are those in which the two jets tend to go
in the virtual photon direction (in the $\gamma p$ centre-of-mass frame) but
separated by a rapidity interval which is small relative to their large
individual rapidities. In this configuration the proton-gluon fusion process
dominates and small values of $x_g$ are sampled. Here $x_g$ is the 
longitudinal fraction of the proton's momentum carried by the interacting gluon.
The process is shown diagrammatically in Fig.\ 1, where the dominant 
structure of the interacting gluon at small $x_g$ is exposed. We also show
the transverse momenta $p_{1T}$, $p_{2T}$ and $k_T$ of 
the outgoing (quark, antiquark) jets and the incoming gluon. 

Dijets can be produced at HERA via direct photons either by photoproduction 
($Q^2 \approx 0$) \cite{JETPHOT,RGRF} or by deep-inelastic electroproduction 
($Q^2$ of O(10 GeV$^2$)). In the former case it is much more difficult to 
extract the direct photon events from the events in which the photon is resolved 
into its constituent partons \cite{AdR}.  In fact
it appears likely that a larger \lq\lq clean\rq\rq\ dijet event sample will
be obtained for electroproduction, and so we study this process.
We are especially interested in the properties of the gluon at small $x_g$.
However the values of $x_g$ that are sampled in any deep inelastic
study exposing the final state are always greater than the Bjorken $x$.
In our case $x_g\approx (1+\hat{s}/Q^2)x$, where $\sqrt{\hat{s}}$ is the
c.m. energy of the produced dijet system.  So the lower the values of
$p_T$, for which the jets can be clearly identified, the better.

In principle, the calculation of the dijet production cross section requires
integration over the complete phase space of the transverse momenta $k_T$
of the gluon. In the conventional approach we (i) restrict the integration
over $k_T^2$ to the region $k_T^2 \ll p_{iT}^2$ and (ii) let 
$k_T^2 = 0$
in the calculation of the subprocess cross section $\hat{\sigma} (\gamma g
\rightarrow q \overline{q})$, i.e. we evaluate $\hat{\sigma}$ with the gluon 
on-mass-shell. As a result we get the familiar factorization formula, which
symbolically is of the form 
\be
x_g g(x_g,\mu^2) \otimes \hat{\sigma} (\gamma g \rightarrow q \overline{q},
\mu^2)
\ee
where the scale $\mu^2 \sim p_{iT}^2$. In this way we can probe the 
conventional 
gluon distribution, $g$, of the proton. However, at small $x_g$, the 
strong-ordering approximation is no longer applicable -- we must
keep the full $k_T^2$ dependence of $\hat{\sigma}$ and integrate over the full
$k_T^2$ phase space. As a consequence we must work in terms of the gluon
distribution $f(x_g, k_T^2, \mu^2)$ unintegrated over $k_T^2$ 
\cite{CATANI1,CATANI2,CELLIS}, that is
\be
x_g g (x_g,\mu^2) = \int^{\mu^2} \frac{dk^2_T}{k^2_T} f(x_g, k_T^2, \mu^2).
\ee

At small $x_g$ the function $f$ satisfies the BFKL equation \cite{BFKL},
\be
\frac{\partial f(x_g, k_T^2)}{\partial {\rm log}(1/x_g)} = 
\frac{3\alpha_s}{\pi} 
k_T^2 \int \frac{dk_T^{\prime 2}}{k_T^{\prime 2}} 
\left[ \frac{f(x_g,k_T^{\prime 2}) - 
f(x_g, k_T^2)}{|k_T^{\prime 2} - k_T^2|} 
+ \frac{f(x_g, k_T^2)}{(4k_T^{\prime 4} + k_T^4)^{1\over 2}} \right] , 
\ee
which effectively resums the large leading log(1/$x_g$) contributions
which arise from the sum of the gluon emission diagrams of the type shown in 
Fig. 1 together 
with the virtual corrections. Note that, at small $x_g$, 
the gluon distribution $f$ becomes independent of the scale $\mu^2$. Two 
characteristic features of the solution of the BFKL equation are (i) the 
leading small $x_g$ behaviour of the form
\be
f(x_g,k_T^2) \sim x_g^{-\lambda}
\ee
where $\lambda \approx 0.5$, and (ii) a \lq\lq diffusion\rq\rq\ in $k_{\rm T}$
with decreasing $x_g$ which arises from the \lq\lq random-walk\rq\rq\ in 
the $k_T$ of the gluons as we proceed along the gluon chain \cite{BL,AKMS2}. The 
first 
property gives rise to a growth in dijet production with decreasing $x_g$.
However this behaviour can be mimicked by eq.\ (1) with a conventional 
gluon which evolves from a singular distribution at some starting scale,
that is $x_g g \sim x_g^{-\lambda}$. The singular small $x_g$ behaviour is 
stable to the evolution. The second property is more unique to BFKL dynamics.
The diffusion in $k_T$ manifests itself in a weakening of the
back-to-back azimuthal correlation of the two outgoing jets of transverse
momenta $p_{1T}$ and $p_{2T}$. As $x$ (and hence $x_g$) decrease, 
larger $k_T$'s
are sampled and broader azimuthal distributions are expected. On the other hand 
the strong-ordering of the transverse momenta of conventional dijet production
leads to a narrow distribution about the back-to-back jet configuration.
Thus, in principle, a measurement of the azimuthal distribution offers
a direct determination of the $k_T$ dependence of the gluon distribution
$f(x_g,k_T^2)$.  In practice the situation is not so clear.
The azimuthal distribution will also be broadened by higher-order
conventional QCD effects.  To see whether these will mask the
BFKL signal we therefore also compute the azimuthal distribution
resulting from the emission of a third ``hard'' QCD jet.

  We begin by using BFKL dynamics at small $x_g$ to
calculate the differential dijet cross section as a function of $\phi$,
the azimuthal angle between the transverse momenta ${\bf p_{1T}}$ and
${\bf p_{2T}}$ of the two jets. That is we evaluate
\be
{d\sigma \over dx dQ^2 d\phi} = {4\pi\alpha^2 \over Q^4 x} \Biggl\lbrack
(1 - y + {y^2 \over 2}) {dF_T (x, Q^2, \phi) \over d\phi}  + (1 - y)
{d F_L (x, Q^2, \phi) \over d\phi} \Biggr\rbrack
\ee
where, as usual, the deep inelastic variables $Q^2 = -q^2$, $x = Q^2 / 2p.q$
and $y = p.q / p.p_e$ where $p_e$, $p$ and $q$ are the four momenta of the
incident electron, proton and virtual photon respectively, see Fig 1. The
differential structure functions $dF_i/d\phi$ can be computed from the 
$k_T$-factorization formula, which is shown symbolically in 
Fig. 1. It is convenient to express the jet four-momenta in terms of 
Sudakov variables 
\begin{eqnarray} 
p_1 & = & (1-\beta)q^{\prime} + \alpha_1p + {\bf p_{1T}} \nonumber \\ 
p_2 & = & \beta q^{\prime} + \alpha_2 p + {\bf p_{2T}} 
\end{eqnarray} 
where $q^{\prime} = q + xp$ and $p$ are basic lightlike momenta.  Thus, since 
the jets are on-mass-shell, we have 
\begin{equation} 
\alpha_1 \; = \; \left( \frac{p^2_{1T}+m^2_q}{(1-\beta)Q^2} \right)x, 
\hspace*{1cm} \alpha_2 \; = \; \left( \frac{p^2_{2T}+m^2_q}{\beta Q^2} \right) x 
\end{equation} 
where $m_q$ is the mass of the quark.  The Bjorken variable $x \approx 
Q^2/2p.q^{\prime}$.  The factorization formula for the differential structure 
functions is 
\begin{equation} 
\frac{dF_i(x,Q^2,\phi)}{d\phi} \; = \; \sum_q \int^1_0 d\beta \int 
\frac{dp^2_{1T}dp^2_{2T}}{k^4_T} f(x_g,k^2_T) {\cal 
F}^q_i(\beta,p^2_{1T},p^2_{2T},\phi) 
\end{equation} 
with $i = T,L$ and where the \lq\lq factorization" variables 
\begin{eqnarray} 
x_g & = & x + \alpha_1 + \alpha_2 \hspace*{.5cm} \left(\gapproxeq [1 + 
4p^2_{iT}/Q^2]x\right) \nonumber \\ 
k^2_T & = & p^2_{1T} + p^2_{2T} + 2p_{1T}p_{2T}{\rm cos}\phi . 
\end{eqnarray} 
The functions ${\cal F}_i^q$ which describe virtual photon-virtual gluon fusion, 
$\gamma g \rightarrow q\bar{q}$, at small $x_g$ are \cite{AKMS1} 
\begin{equation} 
{\cal F}^q_T(\beta,p^2_{1T},p^2_{2T},\phi) = e^2_q \frac{Q^2}{8\pi^2} 
\alpha_s[\beta^2 + (1-\beta)^2] \left\{ \frac{p^2_{1T}+m^2_q}{D^2_1} + 
\frac{p^2_{2T}+m^2_q}{D^2_2} + 2 \frac{p_{1T}p_{2T}{\rm cos}\phi - 
m^2_q}{D_1D_2} \right\} 
\end{equation} 

\begin{equation} 
{\cal F}^q_L(\beta,p^2_{1T},p^2_{2T},\phi) = e^2_q \frac{Q^4}{2\pi^2} \alpha_s 
\beta^2(1-\beta)^2 \left\{ \frac{1}{D^2_1} + \frac{1}{D^2_2} - \frac{2}{D_1D_2} 
\right\}  
\end{equation} 
where $e_q$ is the charge of the quark $q$ and where the denominators 
\begin{equation} 
D_i \; = \; p^2_{iT} + m^2_q + \beta(1-\beta)Q^2 . 
\end{equation} 
The first two terms in \{...\} in (10) and (11) correspond to quark box 
contributions with $p_1$ being first a quark and then an antiquark jet, whereas 
the third term is the \lq\lq crossed-box" interference term.   Note
that the apparent divergence of the integral in (8) at $k^2_T=0$ (that
is at $\phi =\pi$) is cancelled by the zeros of the functions $f$ and
${\cal F}_{T,L}$, see (10) and (11).

  The only unknown in the determination of the differential cross
  section for deep-inelastic dijet production,
$d\sigma/dxdQ^2d\phi$, is the gluon distribution $f(x_g,k_T^2)$ that
enters in (8).  We calculate $f$ by solving the differential form
of the BFKL equation, (3), using knowledge of the gluon at
$x_g=10^{-2}$, as described in ref.\cite{AKMS2}.  The normalisation (though
not the $x_g$ behaviour) of $f$ is dependent on the treatment
of the infrared $k_T$ region.  However, the gluon 
distribution $f$ can also be used
to predict the behaviour of the structure function $F_2$ 
itself, via the inclusive form of the $k_T$-factorization formula (8),
see ref.\cite{AKMS2}.  Thus  we can fix the infrared parameter in the 
determination of $f$ so as to reproduce the low $x$ measurements of
$F_2$ at HERA\cite{HERA}.  The result is shown by continuous curves
in Fig.2, and corresponds
to the choice $k^2_a = 2$ GeV$^2$.  
We see that an excellent
description of $F_2$ is obtained.   Now that $f$ has 
been fully specified in this way,
we should be able to predict the azimuthal distribution reliably,
provided, of course, that we sample sufficiently
small values of $x_g$ for the BFKL solution to be appropriate.

In Fig.\ 3(a) we show $dF_T/d\phi$ calculated from
the $k_T$-factorization formula, (8), for 
deep-inelastic dijet events with $Q^2 = 10$ GeV$^2$ for three different values 
of $x$.  Each jet is required to have transverse energy squared, $E_T^2$, 
greater than 10 GeV$^2$.  
 We notice a rapid increase in the dijet rate with decreasing $x$, 
and a weakening of the azimuthal back-to-back correlation.  This last 
observation is more evident from Fig.\ 3(b) which shows the same distributions 
normalized to a common maximum value at $\phi = \pi$.  The broadening with 
decreasing $x$ is a manifestation of the diffusion in $k_T$ which is 
characteristic of BFKL dynamics.  The detailed shape in the region of $\phi 
\approx \pi$ may not be reliable, since it corresponds to small values of the 
transverse momentum $k_T$, of the gluon (see Fig.\ 1).  Moreover hadronization 
effects will influence the distribution in this region.  Rather we should study 
the normalised distribution, $dF_T/d\phi$, away from $\phi = \pi$, say outside 
the interval 180 $\pm$ 20 degrees.   

  The characteristic BFKL behaviour of the solution $f(x,k_T^2)$
of (3) only has a chance to set in for  $x \lapproxeq 10^{-3}$.
The precocious onset of this leading log$(1/x)$ behaviour does
indeed appear compatible with the striking rise of $F_2$ with
decreasing $x$ that has been observed at HERA, see Fig.2.
However this is not conclusive evidence of BFKL dynamics.
The $F_2$ data can equally well be described by Altarelli-Parisi
(or GLAP) evolution.  For example the dashed curves in Fig.2
are the description obtained from a recent global structure
function analysis based on (next-to-leading order) GLAP
evolution from ``singular'' parton distributions at $Q^2=4$
GeV$^2$\cite{MRSA}.  To distinguish BFKL dynamics from conventional
QCD we must look into properties of the final state at
small $x$, such as the weakening of the back-to-back
correlations in the dijet events.  However there is a price
to pay.  The BFKL dynamics is sampled at larger ``$x$'' 
for final state processes than is the case for
the inclusive $F_2$ measurement.  In our example of dijets
with $E_T^2$(jet) $> 10$ GeV$^2$ and $Q^2 = 10$ GeV$^2$, we 
see from (9) that we sample values of $x_g 
\gapproxeq 5x$.  Thus if $f(x_g,k_T^2)$ assumes a characteristic
BFKL behaviour for $x_g \lapproxeq 10^{-3}$, then  we
anticipate that the broadening of the $\phi$ distribution,
with decreasing $x$, will only be relevant in the region $x \lapproxeq$ few 
$\times 10^{-4}$. This is near the limit of the region which is at present 
accessible at HERA.  In Fig.4 we therefore compare the dijet azimuthal 
distribution for $x = 2 \times 10^{-4}$ with that for $x = 10^{-3}$, that is two 
values of $x$ which are appropriate for HERA.

As mentioned above, the
weakening of the back-to-back azimuthal correlations can also be obtained in 
a more conventional way from fixed-order QCD effects, in
particular from 3+1 jet 
production.  As usual the +1 refers to the 
jet associated with the remnants of the proton.  Part of the 3+1 jet production 
is, of course, already included in the calculation based on BFKL dynamics, since 
the absence of strong-ordering in $k_T$ means that the gluonic ladder contains 
additional (gluon) jets.  Before drawing final conclusions we must therefore 
compare our BFKL dijet
 predictions with the azimuthal distribution coming from
 conventional 3+1 jet production.  To calculate the latter we use
  the PROJET Monte Carlo \cite{GRAUDENZ}. We require two of the jets
to have   
$E_T^2$(jet $1,2) > 10$ GeV$^2$ and the third to have
$E_T^2$(jet $3) < 10$ GeV$^2$,  chosen so that two jets are
``visible'' and the third (which may be either a gluon or a
quark) is relatively ``soft''.  The results are shown by the
histograms superimposed on Fig.4.  We have checked that the PROJET
predictions in the region $|\phi-\pi |\gapproxeq 20^0$ 
are not sensitive to a reasonable variation of the 
cut-off, $y_{ij}\equiv s_{ij}/W^2 > y_0$, that is used to
regulate the infrared singularities.  

We see from Fig.4 that the appearance of dijet
events in the ``tails'' of the azimuthal distribution, that is at
angles such that $|\phi-\pi |\gapproxeq 45^0$, at the predicted rate,
will be a distinctive signal for BFKL dynamics.
Nearer the back-to-back configuration the fixed-order QCD processes
swamp the BFKL effect.   Of course, since we work at the parton
level and ignore the experimental problems of jet identification,
only the BFKL signal can contribute for $|\phi-\pi |>60^0$.  If the
cut, $E_T^2$(jet $3) < 10$ GeV$^2$, on the third jet is removed
and the azimuthal distribution is plotted as a function of the
angle, $\phi $, between the two jets with the largest $E_T$'s then
the PROJET prediction is essentially unchanged for
$|\phi-\pi |\approx 20^0$, but is enhanced at larger angles with
a steeper fall off towards the limiting angle $|\phi-\pi |=60^0$.

The deep inelastic variables $(x,Q^2)$ and jet cuts $(E_T>E_{0T})$
have been chosen in an attempt to optimize both the event
rate and the BFKL signal at HERA.  Clearly if we were able to go to
smaller $x$ (or, to be precise, smaller $x_g$) then
the BFKL effect would be more pronounced, see Fig.3.
Now the observable deep inelastic 
region at HERA lies in the domain $Q^2/x\lapproxeq 
10^5$ GeV$^2$.  On the other hand, we see from (9) that 
$x_g\gapproxeq x+4(x/Q^2)E_{0T}^2$, when  two
jets with $E_T > E_{0T}$ are recorded.  A higher jet threshold, 
$E_{0T}$, has the advantage
that the fixed-order QCD contribution is suppressed relative to
the BFKL signal, but (since $Q^2/x$ is bounded) then a higher
$x_g$ is sampled.  The lower plot in Fig.4 shows the results for
$Q^2=E^2_{0T}=25$ GeV$^2$ and $x=5\times 10^{-4}$, for which
$x_g$ is increased by a factor of 2.5 in comparison to that
sampled for $Q^2=E^2_{0T}=10$ GeV$^2$ and $x=2\times 10^{-4}$.

To sum up, the main aim of our paper is to quantify the observation that BFKL 
dynamics weakens the azimuthal correlation of the $q,\bar{q}$ jets produced in 
small $x$ deep-inelastic scattering via the photon-gluon fusion mechanism.  We 
are able to obtain an absolute prediction since the parameter which specifies 
the infrared contribution to the BFKL equation is chosen such that the 
 measurements of $F_2$ at HERA are reproduced.  (That a physically reasonable 
choice of a single infrared parameter suffices to describe the observed small 
$x$ behaviour of $F_2$ is a far from trivial test of BFKL dynamics.)  For the 
dijet events we find a substantial broadening of the azimuthal distribution and 
an increase of this broadening with decreasing $x$.  However, at HERA
energies, we find that the fixed-order QCD contribution from 3+1 jet production
 exceeds the BFKL signal   near the back-to-back configuration.
Nevertheless at sufficiently large values of $|\phi -\pi |$ 
BFKL dynamics dominates
and give rise to a distinctive ``tail'' to the
azimuthal distribution at an observable rate.  In this way dijet production 
at HERA offers an opportunity to study the $k_T$ dependence of the
gluon distribution of the proton.

\vspace*{1cm}

\noindent {\bf Acknowledgements}
  
\vspace*{1cm}

  We thank Nigel Glover, Genya Levin, Al Mueller,
  Albert De Roeck and Peter Sutton for valuable
  discussions.  This work has been supported in part by the Polish
  KBN grant 2 P302 062 04, by the UK PPA Resesarch Council and the
  EU under contract no. CHRX-CT92-0004.  One of us (AJA) thanks the
  Polish KBN - British Council collaborative research programme for
  support.

\newpage       

\noindent {\bf Figure Captions} 

\begin{itemize} 

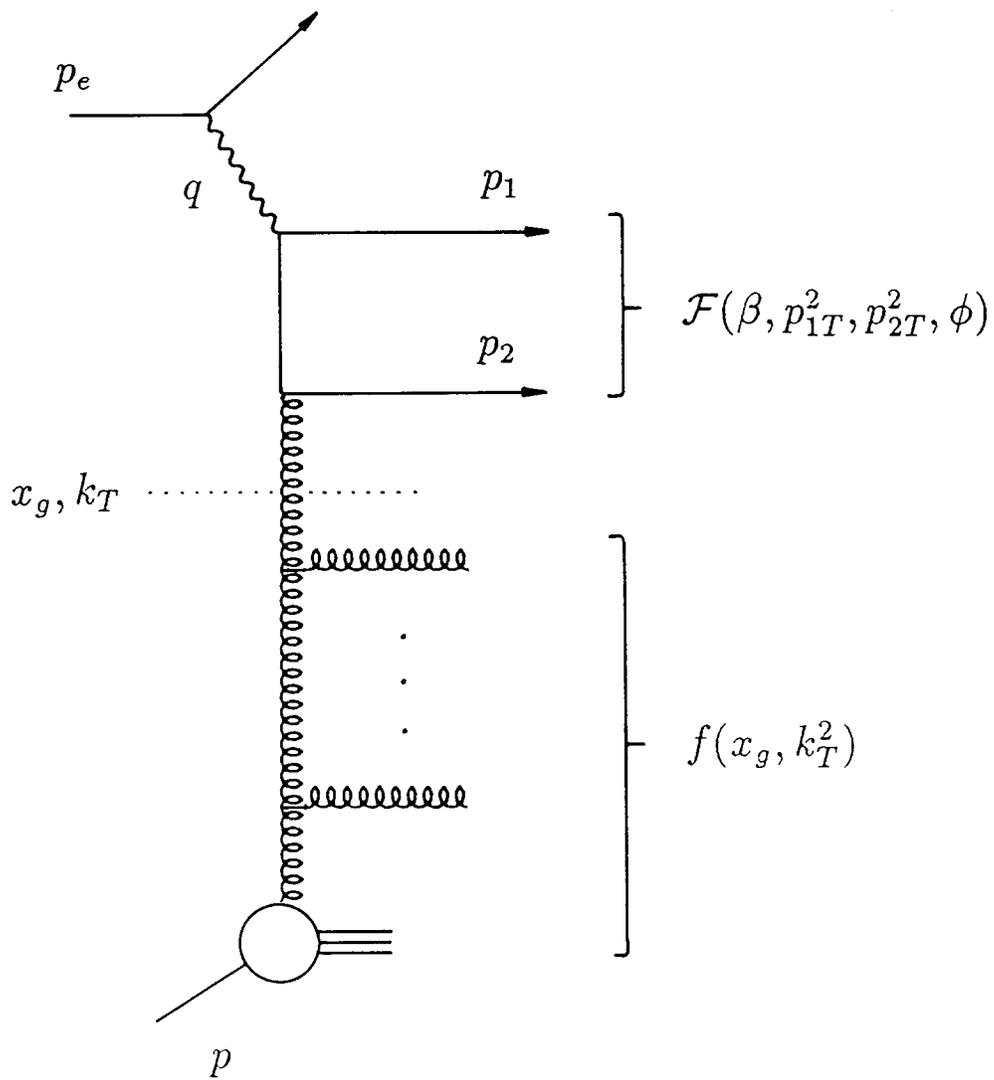
\item[Fig.\ 1] A diagrammatic representation of dijet production by photon-gluon 
fusion, $\gamma g \rightarrow q\bar{q}$, at small $x_g$.  The function $f$ is 
the (unintegrated) gluon distribution of the proton.  The cross section is given 
by the $k_T$-factorization formula (8), which has the symbolic form $\sigma = f 
\otimes {\cal F}$.  ${\cal F}$ denotes the quark box (and crossed box) 
contribution.

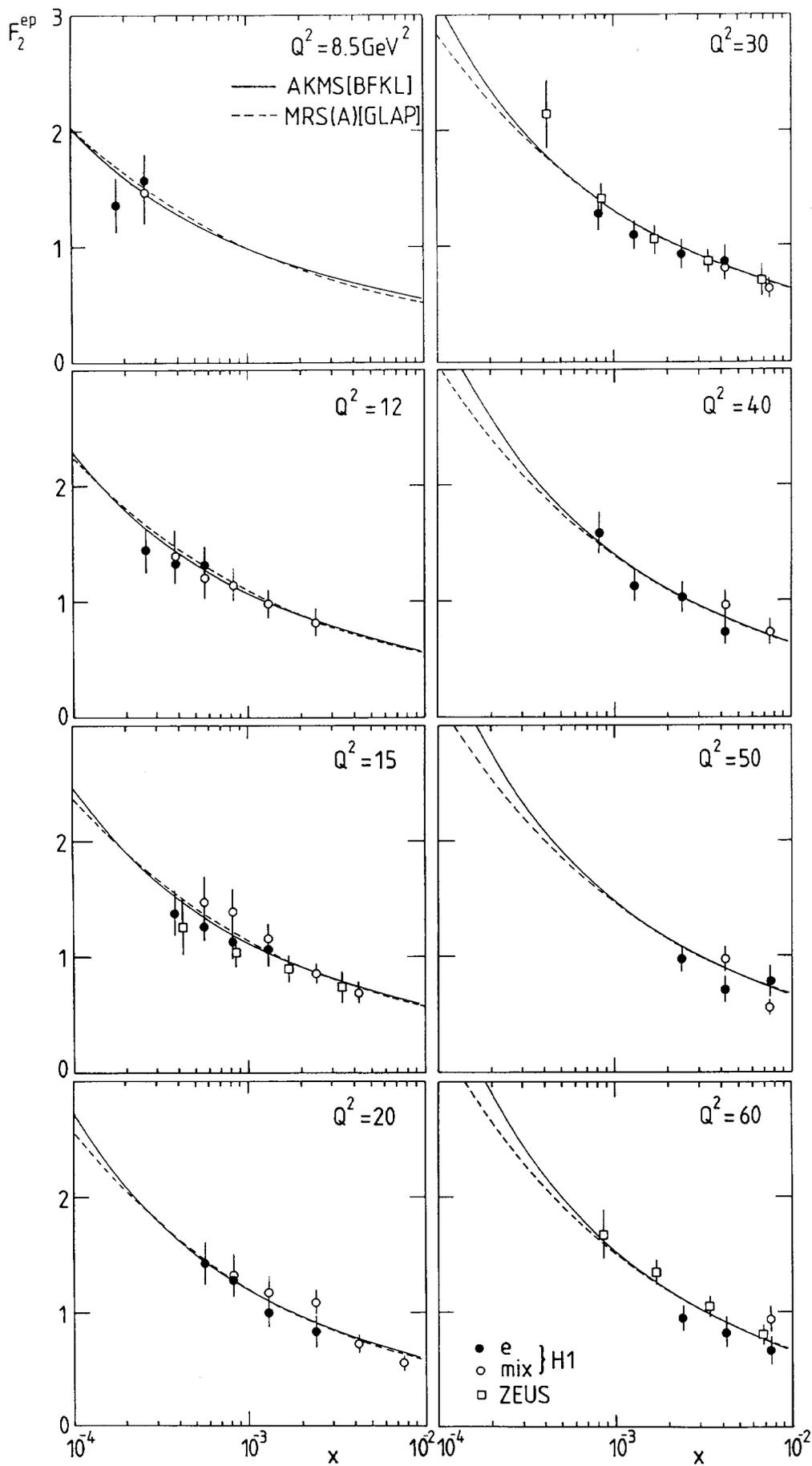
\item[Fig.\ 2] The measurements of $F_2$ at HERA (preliminary data
from the 1993 run\cite{HERA}) shown together with the BFKL
description\cite{AKMS2} (continuous curves) and the MRS(A)
parton analysis fit\cite{MRSA} (dashed curves).   The measurements
of the H1 collaboration at $Q^2=65$ GeV$^2$ are shown on the
$Q^2=60$ GeV$^2$ plot.

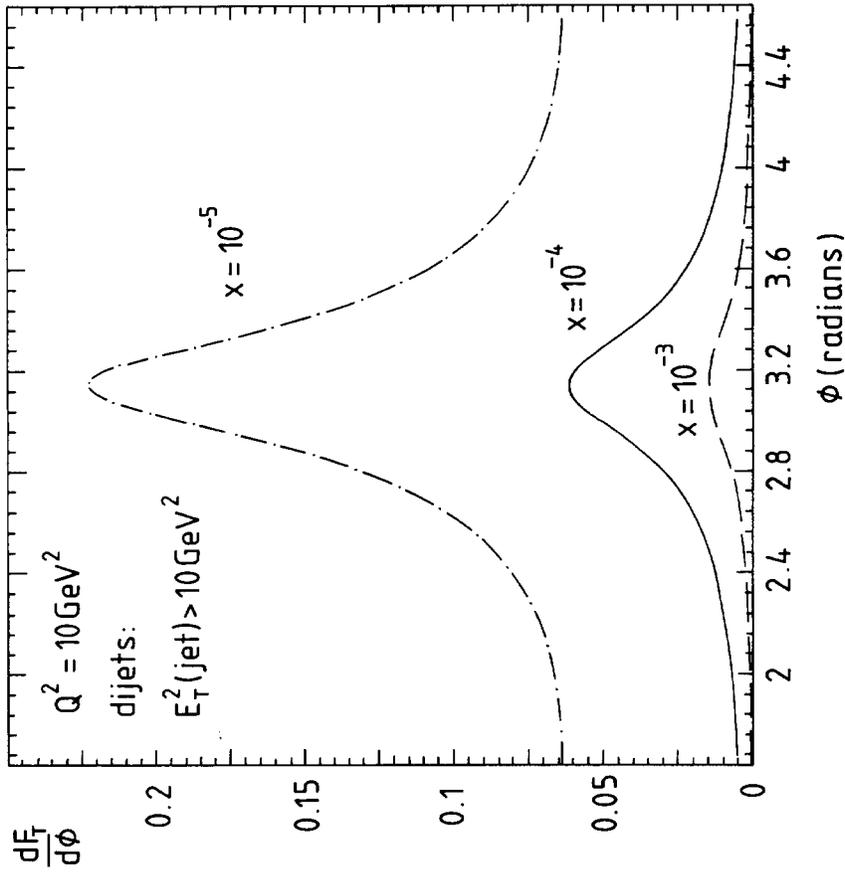
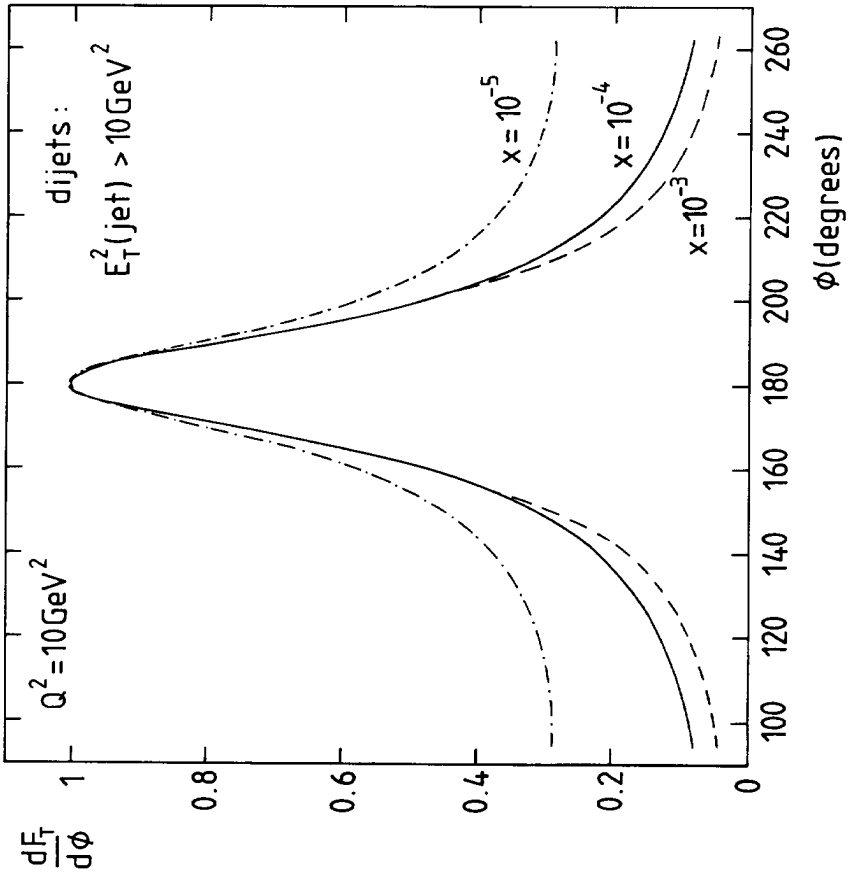
\item[Fig.\ 3] (a) The distribution $dF_T/d\phi$ predicted by the
BFKL $k_T$-factorization formula, (8), for 
deep-inelastic dijet events with $Q^2 = 10$ GeV$^2$ and $E^2_T$(jet) $> 10$ 
GeV$^2$.  (b) The distributions normalized to a common maximum.

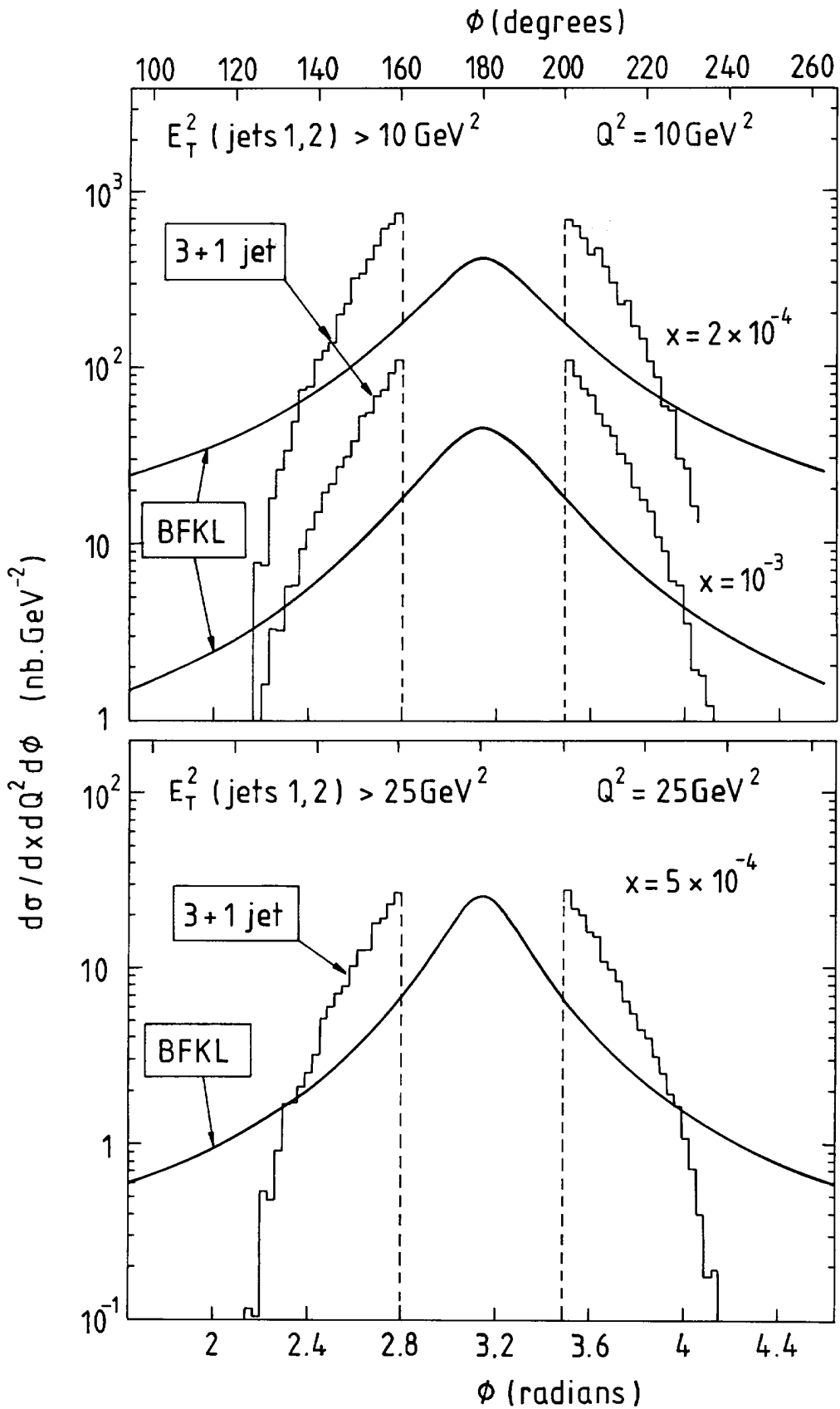
\item[Fig.\ 4] The curves are the differential cross section for dijet 
production predicted by BFKL dynamics, 
(5)-(12), whereas the histograms correspond to 3+1 jet 
production as determined by the PROJET Monte Carlo using MRS(A) partons
\cite{MRSA}.  In the first case, the broadening of the azimuthal
distribution arises from the $k_T$ dependence of the gluon
distribution $f(x_g,k_T^2)$ found by solving the BFKL equation, (3).  
For 3+1 jet production we assume that the third 
jet has $E_T^2<10(25)$ GeV$^2$ in the upper (lower) plot.
\end{itemize} 

\end{document}